\documentclass[aps,prl,twocolumn,showpacs,superscriptaddress]{revtex4}
\usepackage{graphicx}
\usepackage{bm}
\newcommand{\beq}{\begin{equation}}
\newcommand{\eeq}{\end{equation}}

\begin{document}

\title{\boldmath Hall coefficient in heavy fermion metals}

\author{V.R. Shaginyan \footnote{E--mail:
vrshag@thd.pnpi.spb.ru}}
\affiliation{Petersburg Nuclear Physics Institute, RAS,
Gatchina, 188300, Russia}
\author{K.G. Popov}
\affiliation{Komi Science Center,
Ural Division, RAS,
3a, Chernova str.
Syktyvkar, 167982, Russia}
\author{S.A. Artamonov}
\affiliation{Petersburg Nuclear Physics Institute, RAS,
Gatchina, 188300, Russia}


\begin{abstract}

Experimental studies of the antiferromagnetic (AF) heavy
fermion metal $\rm YbRh_2Si_2$ in a magnetic field $B$ indicate
the presence of a jump in the Hall coefficient at a magnetic-field
tuned quantum state in the zero temperature limit.
This quantum state occurs at $B\geq B_{c0}$ and induces the jump  even
though the change of the magnetic field at $B=B_{c0}$ is infinitesimal.
We investigate this by using the model of heavy electron liquid with
the fermion condensate. Within this model the jump takes place when the
magnetic field reaches the critical value $B_{c0}$ at which the
ordering temperature $T_N(B=B_{c0})$ of the AF transition vanishes. We
show that at $B\to B_{c0}$, this second order AF phase transition
becomes the first order one, making the corresponding quantum and
thermal critical fluctuations vanish at the jump. At $T\to0$ and
$B=B_{c0}$, the Gr\"uneisen ratio as a function of temperature $T$
diverges. We demonstrate that both the divergence and the jump are
determined by the specific low temperature behavior of the entropy
$S(T)\propto S_0+a\sqrt{T}+bT$ with $S_0$, $a$ and $b$ are temperature
independent constants.

\end{abstract}


\maketitle

The most outstanding puzzle of heavy fermion (HF) metals
is what determines their universal behavior which drastically differs
from the behavior of ordinary metals. It is wide accepted that the
fundamental physics observed in the HF metals is controlled by quantum
phase transitions. A quantum phase    transition is  driven by control
parameters such as composition, pressure, number density $x$ of
electrons (holes), magnetic field $B$, etc, and takes place at a
quantum critical point (QCP) when the temperature $T=0$. In the case of
conventional quantum phase transitions (CQPT) the physics is dominated
by thermal and quantum fluctuations near CQP. This critical state is
characterized by the absence of quasiparticles. It is believed
that the absence of quasiparticle-like excitations is the main
cause of the non-Fermi liquid (NFL) behavior, see e.g. \cite{voj}.
However, theories based on CQPT fail to explain the experimental
observations of the universal behavior related to the divergence of the
effective mass $M^*$ at the magnetic field tuned QCP, the specific
behavior of the spin susceptibility, its scaling properties, etc.

It is possible to explain the observed universal behavior
of the HF metals on the basis of the
fermion condensation quantum phase transition (FCQPT) which takes
place at $x=x_{FC}$ and allows the existence of the Landau
quasiparticles down to the lowest temperatures \cite{shag2}. It is the
quasiparticles which define the universal behavior of the HF metals at
low temperatures \cite{shag2,ckhz}. In contrast to the conventional
Landau quasiparticles, these are characterized by the effective mass
which  strongly depends on temperature $T$, applied magnetic field
$B$ and the number density $x$ of the heavy electron liquid of HF
metal. Thus, we come back again to the key role of the of the
effective mass.

On the other hand, it is plausible
to probe the other properties of the
heavy electron liquid which are not directly determined by the
effective mass.  Behind the point of FCPT when $x<x_{FC}$, the heavy
electron liquid possesses unique features directly determined by its
quasiparticle distribution function $n_0({\bf p})$ formed by the
presence of the fermion condensate (FC) \cite{ks}. Therefore, the
function $n_0({\bf p})$ drastically differs from the quasiparticle
distribution function of a typical Landau Fermi liquid (LFL)
\cite{lanl1}. For example, it was predicted that at low temperatures
the  tunneling differential conductivity between  HF metal with FC and
a simple metallic point can be noticeably dissymmetrical with respect
to the change of voltage bias \cite{tun}. As we shall see below, the
magnetic field dependence of the Hall coefficient $R_H(B)$ can also
provide information about electronic systems with FC.

Recent experiments have shown that the Hall coefficient in
the antiferromagnetic (AF)  HF metal $\rm YbRh_2Si_2$ in a magnetic
field $B$ undergoes a jump in the zero temperature limit upon
magnetic-field tuning the metal from  AF to a paramagnetic state
\cite{pash}. At some critical value $B_{c0}$, the magnetic field $B$
induces the jump  even though the change of the magnetic field at at
the critical value $B_{c0}$ is infinitesimal.

In this letter, we show that the abrupt change in the Hall coefficient
is determined by the presence of FC and investigate this jump by using
the model of the heavy electron liquid with FC which is represented by
an uniform electron liquid near FCQPT. Within this model the jump takes
place when magnetic field reaches the critical value $B_{c0}$ at which
the N\'eel temperature $T_N(B=B_{c0})$ of the AF transition vanishes.
At some temperature $T_{crit}$ when $B\to B_{c0}$, this second order AF
phase transition becomes the first order one, making the corresponding
quantum and thermal critical fluctuations vanish at the point where
$T_N(B=B_{c0})\to0$. At $T\to0$ and $B=B_{c0}$, the Gr\"uneisen  ratio
$\rm{\Gamma}(T)=\alpha(T)/C(T)$ as a function of temperature $T$
diverges. Here, $\alpha(T)$ is the thermal expansion coefficient and
$C(T)$ is the specific heat. We show that both the divergence and the
jump are determined by the specific low temperature behavior of the
entropy $S(T)\simeq S_0+a\sqrt{T/T_f}+bT/T_f$ with $S_0$, $a$ and $b$
being temperature independent constants, and $T_f$ is the temperature
at which the influence of FC vanishes.

To study the universal behavior of the HF metals at low temperatures,
we use the heavy electron liquid model  in order to get rid of the
specific peculiarities of a HF metal. It is possible since we consider
processes related to the power-low divergences  of the corresponding
physical quantities. These divergences are determined by small momenta
transferred as compared to momenta of the order of the reciprocal
lattice, therefore, the contribution coming from the lattice can be
ignored. On the other hand, we can simply use the common concept of the
applicability of the LFL theory when describing electronic properties
of metals \cite{lanl1}. Thus, we may safely ignore the complications
due to the anisotropy of the lattice regarding the medium as the
homogeneous heavy electron isotropic liquid.

At first, we briefly describe the heavy electron liquid with FC.
Dealing with FCQPT, we have to put $T=0$.
In that case, the ground state energy $E_{gs}$
of a system in the superconducting state is given by the BSC
theory formula \beq E_{gs}[\kappa({\bf p})]=E[n({\bf p})]+
E_{sc}[\kappa({\bf p})], \eeq where the occupation numbers $n({\bf
p})$ are connected to the factors $v({\bf p})$, $u({\bf p})$
and the order parameter $\kappa({\bf p})$ $$ n({\bf
p})=v^2({\bf p});\,\,\, v^2({\bf p})+u^2({\bf p})=1;$$
\beq \kappa({\bf p})=v({\bf p})u({\bf p})
=\sqrt{n({\bf p})(1-n({\bf p}))}.\eeq
The second term $E_{sc}[\kappa_p]$
on the right hand side of Eq. (1) is defined by the
superconducting contribution which in the simplest case of the
weak coupling regime is of the form \beq E_{sc}[\kappa_p]= \lambda\int
V_{pp}({\bf p}_1,{\bf p}_2)\kappa({\bf p}_1) \kappa^*({\bf p}_2)
\frac{d{\bf p}_1d{\bf p}_2}{(2\pi)^4}, \eeq
where $\lambda V_{pp} ({\bf p}, {\bf p}_1)$ is the pairing interaction.
Varying $E_{gs}$ given by Eq. (1) with respect to $v({\bf
p})$ one finds \beq \varepsilon({\bf p})-\mu=\Delta({\bf
p})\frac{1-2v^2({\bf p})}{2\kappa({\bf p})}.\eeq Here
$\varepsilon({\bf p})$ is defined by the Landau equation
$\delta E[n({\bf p})]/\delta n({\bf p})=\varepsilon({\bf p}),$
$\mu$ is chemical potential, and the gap \beq
\Delta({\bf p})=-\lambda\int V_{pp} ({\bf p}, {\bf p}_1)
\sqrt{n({\bf p}_1)(1-n({\bf p}_1))} \frac{d{\bf p}_1}{4\pi^2}.\eeq
If $\lambda \to 0$, then
$\Delta({\bf p})\to 0$, and Eq. (4) reduces to the equation \beq
\frac{\delta E[n({\bf p})]}{\delta n({\bf
p})}-\mu=
\varepsilon({\bf p})-\mu=0,\: {\rm {if}}\,\,\, \kappa({\bf p})\neq 0. \eeq
As a result, at $x<x_{FC}$, the
function $n({\bf p})$ is determined by the standard equation
to search the minimum of functional $E[n({\bf p})]$ \cite{dkss,vsl}.
Equation (6) determines the quasiparticle distribution function
$n_0({\bf p})$ which delivers the minimum  value to the ground
state energy $E$. The function $n_0({\bf p})$ being the signature of
the new state of quantum liquids \cite{vol} does not coincide with the
step function in the region $(p_f-p_i)$ where $\kappa({\bf p})\neq 0$,
so that $0<n_0({\bf p})<1$ and $p_i<p_F<p_f$, with
$p_F=(3\pi^2x)^{1/3}$ is the Fermi momentum. We note the remarkable
peculiarity of FCQPT at $T=0$: this transition is related to
spontaneous breaking of gauge symmetry, when the superconducting
order parameter $\kappa({\bf p})=\sqrt{n_0({\bf p})(1-n_0({\bf p}))}$
has a nonzero value over the
region occupied by the fermion condensate,
with the entropy $S=0$ \cite{vsl,shag2}, while the gap $\Delta({\bf p})$
vanishes provided that $\lambda \to0$ \cite{dkss,vsl}.
Thus the state with FC cannot exist at
any finite temperatures and driven by the parameter $x$: at
$x>x_{FC}$ the system is on the disordered side of FCQPT; at
$x=x_{FC}$, Eq. (6) possesses the non-trivial solutions $n_0({\bf
p})$ with $p_i=p_F=p_f$; at $x<x_{FC}$, the system is on the
ordered side \cite{shag2}.

At finite temperatures $0<T\ll T_f$, the function $n_0({\bf p})$
determines the entropy $S_{NFL}(T)$ of the heavy electron
liquid in its NFL state $$
S_{NFL}[n(p)]=-2\int [ n({\bf p},T)\ln n({\bf p},T)+(1-n({\bf p},T))$$
\beq\times\ln (1-n({\bf p},T))] \frac{d{\bf p}}{(2\pi )^3},\eeq
with $T_f$ being the temperature at which the influence of FC vanishes
\cite{dkss,vsl}. Inserting into Eq. (7) the function $n_0({\bf p})$,
one can check that behind the point of FCQPT there is a temperature
independent contribution $S_0(r)\propto(p_f-p_F)\propto|r|$,  where
$r=x_{FC}-x$.  Another specific contribution is related to the spectrum
$\varepsilon ({\bf p})$ which insures the connection between the
dispersionless region $(p_f-p_i)$ occupied by FC and the normal
quasiparticles located at $p<p_i$ and at $p>p_f$, and therefore it is
of the form $\varepsilon ({\bf p})\sim (p-p_f)^2\sim (p_i-p)^2$.  Such
a form of the spectrum can be verified in exactly solvable models for
systems with FC and leads to the contribution of this spectrum to the
specific heat $C\sim\sqrt{T/T_f}$ \cite{ks}.  Thus at $0<T\ll T_f$, the
entropy can be approximated as \beq S_{NFL}(T) \simeq
S_0(r)+a\sqrt{\frac{T}{T_f}}+b\frac{T}{T_f}, \eeq with $a$ and $b$ are
constants. The third term on the right hand side of Eq. (8) comes from
the contribution of the temperature independent part of the spectrum
$\varepsilon({\bf p})$ and gives a relatively small contribution to the
entropy.

The temperature independent term $S_0(r)$ determines the specific NFL
behavior of the system. For example, the thermal expansion coefficient
$\alpha(T)\propto x{\partial (S/x)}/{\partial x}$ determined mainly by
the contribution coming from $S_0(r)$  becomes constant at $T\to0$
\cite{zver}, while the specific heat $C=T\partial S(T)/\partial T\simeq
(a/2)\sqrt{T/T_f})$. As a result, the Gr\"uneisen ratio $\Gamma(T)$
diverges as $\Gamma(T)=\alpha(T)/C(T)\propto \sqrt{T_f/T}$.

We see that  at $0<T\ll T_f$, the heavy electron liquid  behaves
as if it were placed at QCP, in fact it is placed at the quantum
critical line $x<x_{FC}$, that is the critical behavior is observed at
$T\to0$ for all $x\leq x_{FC}$. At $T\to0$, the heavy electron liquid
undergoes a first-order quantum phase transition because the entropy is
not a continuous function: at finite temperatures the entropy is given
by Eq. (8), while $S(T=0)=0$. Therefore, the entropy undergoes a sudden
jump $\delta S=S_0(r)$ in the zero  temperature limit. We make up a
conclusion that due to the first order phase transition, the critical
fluctuations are suppressed at the quantum critical line and the
corresponding divergences, for example the divergence of ${\rm
\Gamma}(T)$,  are determined by the quasiparticles rather than by the
critical fluctuations as one could expect in the case of  CQPT, see
e.g. \cite{voj}. Note that according to the well known inequality,
$\delta Q\leq T\delta S$, the heat $\delta Q$ of the transition from
the ordered phase to the disordered one is equal to zero, because
$\delta Q\leq S_0(r)T\to 0$ at $T\to 0$.

To study the nature of the abrupt change in the Hall coefficient,
we consider the case when the LFL behavior arises by the suppression
of the AF phase upon applying a magnetic field $B$, for example, as it
takes place in the HF metals $\rm YbRh_2Si_2$ and
YbRh$_2$(Si$_{0.95}$Ge$_{0.05}$)$_2$ \cite{geg,geg1}. The AF phase is
represented by the heavy electron LFL, with the entropy vanishing as
$T\to 0$. For magnetic fields exceeding the critical value $B_{c0}$ at
which the N\'eel temperature $T_N(B\to B_{c0})\to 0$ the weakly ordered
AF phase transforms into weakly polarized heavy electron LFL. At $T=0$,
the application of the magnetic field $B$ splits the FC state occupying
the region $(p_f-p_i)$ into the Landau levels and suppresses the
superconducting order parameter $\kappa({\bf p})$ destroying the FC
state.  Such a state is given by the multiconnected Fermi sphere, where
the smooth quasiparticle distribution function $n_0({\bf p})$ in the
$(p_F-p_i)$ range is replaced by a multiconnected distribution.
Therefore the LFL behavior is restored being represented by the weakly
polarized heavy electron LFL and characterized by quasiparticles with
the effective mass $M^*(B)$  \cite{shag2,shag} \beq M^{*}(B)\propto
\frac{1}{\sqrt{B-B_{c0}}}.\eeq At elevated temperatures
$T>T^*(B-B_{c0})\propto \sqrt{B-B_{c0}}$, the NFL state is restored and
the entropy of the heavy electron liquid is given by Eq. (8). This
behavior is displayed in the $T-B$ phase diagram shown in Fig. 1.

\begin{figure}[!ht]
\begin{center}
\includegraphics[width=0.47\textwidth]{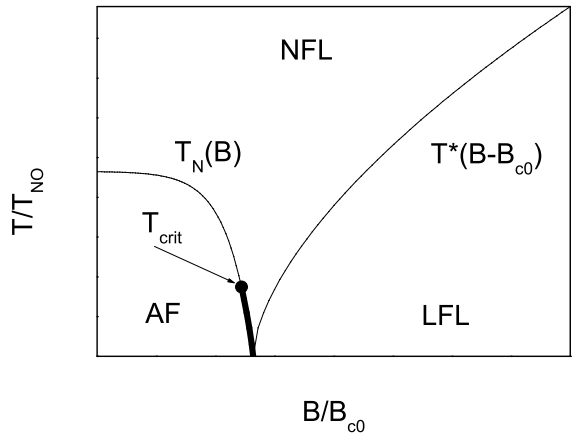}
\end{center}
\caption{$T-B$ phase diagram of the heavy electron liquid. The
$T_N(B)$ curve represents the field dependence of the N\'eel
temperature. Line separating the antiferromagnetic (AF) and the
non-Fermi liquid (NFL) state is a guide to the eye. The black dot at
$T=T_{crit}$ shown by the arrow is the critical temperature at which
the second order AF phase transition becomes the first one. At
$T<T_{crit}$, the thick solid line represents the field dependence of
the N\'eel temperature when the AF phase transition is of the first
order.  The NFL state is characterized by the entropy $S_{NFL}$ given
by Eq. (8). Line separating the NFL state and the weakly polarized
heavy electron Landau Fermi Liquid (LFL) is $T^*(B-B_{c0})\propto
\sqrt{B-B_{c0}}$.} \label{fig1} \end{figure}

In accordance with experimental facts we assume that at relatively
high temperatures $T/T_{NO}\sim 1$ the AF phase transition is of the
second order \cite{geg}. Where $T_{NO}$ is the N\'eel temperature in
the absence of the magnetic field. In that case, the entropy and the
other thermodynamic functions are continuous functions at the
transition temperature $T_N(B)$. This means that the entropy of the AF
phase $S_{AF}(T)$ coincides with the entropy of the NFL state given by
Eq. (8), \beq S_{AF}(T\to T_N(B))=S_{NFL}(T\to T_N(B)).\eeq Since the
AF phase demonstrates the LFL behavior, that is $S_{AF}(T\to 0)\to0$,
Eq. (10) cannot be satisfied at sufficiently low temperatures $T\leq
T_{crit}$ due to the temperature-independent term $S_0(r)$, see Eq.
(8). Thus, the second order AF phase transition becomes the first order
one at $T=T_{crit}$ as it is shown in Fig. 1. At $T=0$, the critical
field $B_{c0}$ at which the AF phase becomes the heavy LFL is
determined by the condition that the ground state energy of the AF
phase coincides with the ground state energy $E[n_0({\bf p})]$ of the
heavy LFL, that is the ground state of the AF phase becomes degenerated
at $B=B_{c0}$. Therefore, the N\'eel temperature $T_N(B\to B_{c0})\to
0$, and the behavior of the effective mass $M^*(B\geq B_{c0})$ is given
by Eq. (9), that is $M^*(B)$  diverges when $B\to B_{c0}$.  We note
that the corresponding quantum and thermal critical fluctuations vanish
at $T<T_{crit}$ because we are dealing with the first order AF  phase
transition. We can also safely conclude that the critical behavior
observed at $T\to0$ and $B\to B_{c0}$ is determined by the
corresponding quasiparticles rather than by the critical fluctuations
accompanying second order phase transitions. When $r\to 0$ the heavy
electron liquid approaches FCQPT from the ordered phase. Obviously,
$T_{crit}\to0$ at the point $r=0$, and we are led to the conclusion
that the N\'eel temperature vanishes at the point when the AF second
order phase transition becomes the first order one. As a result, one
can expect that the contributions coming from the corresponding
critical fluctuations can only lead to the logarithmic corrections  to
the Landau theory of the phase transitions \cite{lanl2}, and the power
low critical behavior is again defined by the corresponding
quasiparticles.

Now we are in position to consider the recently observed jump in the
Hall coefficient at $B\to B_{c0}$ in the zero temperature limit
\cite{pash}.  At $T=0$, the application of the critical magnetic field
$B_{c0}$ suppressing the AF phase (with the Fermi momentum
$p_{AF}\simeq p_F$) restores the LFL with the Fermi momentum $p_f>p_F$.
At $B<B_{c0}$, the ground state energy of the AF phase is lower then
that of the heavy LFL, while at $B>B_{c0}$, we are dealing with the
opposite case, and  the heavy LFL wins the competition.  At $B=B_{c0}$,
both AF and LFL have the same ground state energy being degenerated .
Thus, at $T=0$ and $B=B_{c0}$, the infinitesimal change in the magnetic
field $B$ leads to the finite jump in the Fermi momentum. In response
the Hall coefficient $R_H(B)\propto 1/x$ undergoes the corresponding
sudden jump. Here we have assumed that the low temperature $R_H(B)$ can
be considered as a measure of the Fermi volume and, therefore, as a
measure of the Fermi momentum \cite{pash}{. As a result, we obtain \beq
\frac{R_H(B=B_{c0}-\delta)}{R_H(B=B_{c0}+\delta)}\simeq
1+3\frac{p_f-p_F}{p_F}\simeq 1+d\frac{S_0(r)}{x_{FC}}.
\eeq Here $\delta$ is infinitesimal magnetic field, $S_0(r)/x_{FC}$ is
the entropy per one heavy electron, and $d$ is a constant, $d\sim 5$.
It follows from Eq. (11) that the abrupt change in the Hall coefficient
tends to zero when $r\to0$ and vanishes when the system in question is
on the disordered side of FCQPT.

As an application of the above consideration we study the $T-B$ phase
diagram for the HF metal ${\rm YbRh_2Si_{2}}$ \cite{geg} shown in Fig.
2. The LFL behavior is characterized by the effective mass $M^*(B)$
which diverges as $1/\sqrt{B-B_{c0}}$ \cite{geg}. We can conclude that
Eq. (9) gives good description of this experimental fact, and $M^*(B)$
diverges at the point $B\to B_{c0}$ with $T_N(B=B_{c0})=0$. It is seen
from Fig. 2, that the line separating the LFL state and NFL can be
approximated by the function $c\sqrt{B-B_{c0}}$ with $c$ being a
parameter.  Taking into account that the behavior of YbRh$_2$Si$_{2}$
strongly resembles the behavior of YbRh$_2$(Si$_{0.95}$Ge$_{0.05}$)$_2$
\cite{geg1,cust,geg2}, we can conclude that in the NFL state the
thermal expansion coefficient $\alpha(T)$ does not depend on $T$ and
the Gr\"uneisen ratio as a function of temperature $T$ diverges
\cite{geg1}. We are led to the conclusion that the entropy of the NFL
state is given by Eq. (8). Taking into account that at relatively high
temperatures the AF phase transition is of the second order \cite{geg},
we predict that at lower temperatures this becomes the first order
phase transition.  Then, the described behavior of the Hall coefficient
$R_H(B)$ is in good agreement with experimental facts \cite{pash}.

\begin{figure}[!ht]
\begin{center}
\includegraphics[width=0.47\textwidth]{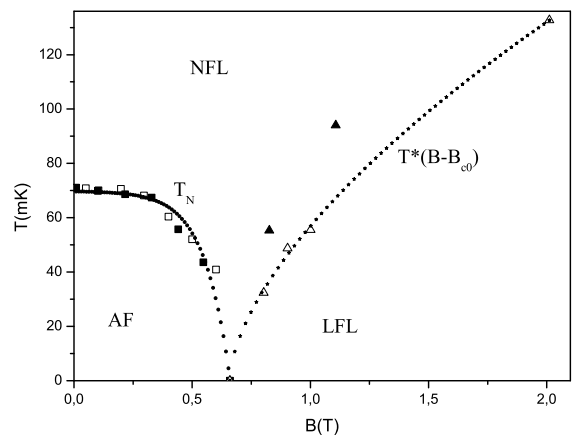}
\end{center}
\caption{$T-B$ phase diagram for ${\rm YbRh_2Si_2}$. The $T_N$ curve
represents the field dependence of the N\'eel temperature. Line
separating the antiferromagnetic (AF) and the non-Fermi liquid (NFL)
state is a guide to the eye.  The NFL state is characterized by the
entropy $S_{NFL}$ given by Eq. (8). Line separating the NFL state and
the Landau Fermi Liquid (LFL) is $T^*(B-B_{c0})=c\sqrt{B-B_{c0}}$, with
$c$ being an adjustable factor.} \label{fig2} \end{figure}

In summary, we have shown that the $T-B$ phase diagram of the heavy
electron liquid with FC is in good agreement with the experimental
$T-B$ phase diagram obtained in measurements on the HF metals $\rm
YbRh_2Si_2$ and YbRh$_2$(Si$_{0.95}$Ge$_{0.05}$)$_2$.  We have also
demonstrated that the abrupt jump in the Hall coefficient $R_H(B)$ is
determined by the presence of FC.  We observed that at decreasing
temperatures $T\leq T_{crit}$, the second order AF phase transition
becomes the first order one, making the corresponding quantum and
thermal critical fluctuations vanish at the jump. Therefore, the abrupt
jump and the divergence of the effective mass taking place at $T_N\to0$
are defined by the behavior of quasiparticles rather than by the
corresponding thermal and quantum critical fluctuations.

This work was supported by Russian Foundation for Basic Research,
Grant No 05-02-16085.


\begin{thebibliography}{99}

\bibitem{voj} M. Vojta, Rep. Prog. Phys. {\bf 66}, 2069 (2003).

\bibitem{shag2}  V.R. Shaginyan, JETP Lett. {\bf 79}, 286 (2004);
V.R. Shaginyan, A.Z. Msezane, and M.Ya. Amusia, Phys. Lett. A {\bf
338}, 393 (2005).

\bibitem{ckhz} J.W. Clark, V.A. Khodel, and M.V. Zverev,
Phys. Rev. B {\bf 71}, 012401 (2005).

\bibitem{ks} V.A. Khodel and V.R. Shaginyan,
JETP Lett. {\bf 51}, 553 (1990); V.A. Khodel, V.R. Shaginyan, and
V.V. Khodel, Phys. Rep. {\bf 249}, 1 (1994).

\bibitem{lanl1}  E.M. Lifshitz and L.P. Pitaevskii,
{\it Statistical Physics,}
Part 2, Butterworth-Heinemann (1999).

\bibitem{tun}  V.R. Shaginyan, JETP Lett. {\bf 81}, 222 (2005).

\bibitem{pash} S. Paschen {\it et al.,} Nature {\bf 432}, 881 (2004).

\bibitem{dkss} J. Dukelsky, V.A. Khodel, P. Schuck, and V.R.
Shaginyan, Z. Phys. {\bf 102}, 245 (1997); V.A. Khodel and V.R.
Shaginyan, Condensed Matter Theories, {\bf 12}, 222 (1997).

\bibitem{vsl} V.R. Shaginyan, Phys. Lett. A {\bf 249}, 237 (1998);
M.Ya. Amusia and V.R. Shaginyan, Phys. Rev. B {\bf 63}, 224507 (2001).

\bibitem{vol} G. E. Volovik, JETP Lett. {\bf 53}, 222 (1991).

\bibitem{zver} M.V. Zverev, V.A. Khodel, V.R. Shaginyan, and M. Baldo,
JETP Lett. {\bf 65}, 863 (1997).

\bibitem{geg} P. Gegenwart {\it et al.,} Phys. Rev. Lett.
{\bf 89}, 056402 (2002).

\bibitem{geg1}  R. K\"uchler {\it et al.,}
Phys. Rev. Lett. {\bf 91}, 066405 (2003).

\bibitem{shag}  Yu.G. Pogorelov and V.R. Shaginyan,
JETP Lett. {\bf 76}, 532 (2002).

\bibitem{lanl2}  E.M. Lifshitz and L.P. Pitaevskii,
{\it Statistical Physics,} vol.5, Pergamon (1980).

\bibitem{cust} J. Custers {\it et al.,} Nature {\bf 424}, 524 (2003).

\bibitem{geg2} P. Gegenwart {\it et al.,}
Phys. Rev. Lett. {\bf 94}, 076402 (2005).

\end{thebibliography}
\end{document}